# A note on the normalization of the momentum eigenfunctions and Dirac delta function

**M. Hage-Hassan**
Université Libanaise, Faculté des Sciences Section (1)
Hadath-Beyrouth

**Abstract**
We determine the generating function of the harmonic oscillator by a new method. Using this generating function we derive the eigenfunctions of the moment p. We find that the normalization of these eigenfunctions is a real and not complex number with phase factor chosen equal one (standard books of quantum mechanics). We prove that the integral of the delta function is equal to one and we derive the oscillator propagator.

## 1. Introduction

Despite the success of quantum mechanics many problems remain unsolved or solved by difficult methods. Among these problems the normalization $N_p$ of the eigenfunctions of the moment which is assumed to be a complex number but « the phase factor is chosen to be equal to unity » [1.p:330, 2.p:54, 3.p:101]. We need also simple methods to determine: the generating function of the basis of the oscillator [3], the integration of delta function and the Feynman propagator of the oscillator [1-3].

We use the generating function method that we develop and that has enough variety of applications [4] to solve these problems. So we show that normalization is real and not arbitrary and that $N_p = \sqrt{1/2\pi\hbar}$ and the calculations of other problems occur without particular difficulty using the Gaussian integral.

In the second part we give a quick review of the wave functions in position and momentum. The third and the fourth part are devoted to a revision of the linear harmonic oscillator and the derivation of the generating function of the oscillator. We devote the sixth part to the derivation of the wave function of the momentum. We treat the properties of Dirac delta function in part sixth. In part seven, we give the derivation of the Feynman propagator of the harmonic oscillator without the use of Mehler's formula [1,2].

## 2. The wave functions in position and momentum

In quantum mechanics the wave function is expressed in terms of coordinate {x} or momentum{p}.
A- In the coordinate representation we write:
$$x, \quad p = \frac{\hbar}{i}\frac{d}{dx}, \quad and \quad [x, p_x] = i\hbar \tag{2.1}$$

Following Dirac the eigenfunctions of the observable $x_o$ can be written:
$$x_o|x\rangle = x|x\rangle, \quad and \quad \langle x'|x_o|x\rangle = x\langle x'\|x\rangle = x'\langle x'\|x\rangle$$

Therefore:  $(x - x')\langle x'\|x\rangle = 0$  (2.2)

The normalization of this space is the Dirac delta function which is defined by:
$$\delta(x - x') = \langle x'\|x\rangle \quad (2.3)$$

B- In moment representation (p, $x = i\hbar d/dp$) the wave function is $p(x) = \langle x\|p\rangle$,

and  $\langle x|p|p\rangle = pp(x), \quad \langle x|\frac{\hbar}{i}\frac{d}{dx}|p\rangle = \frac{\hbar}{i}\frac{d}{dx}p(x)$  (2.4)

The solution of the equation $\frac{\hbar}{i}\frac{d}{dx}p(x) = pp(x)$ is:
$$p(x) = N_p e^{ixp/\hbar} \quad (2.5)$$

We will determine the normalization $N_p$ in part five.
C- We write the unitary operators in the two representations by:
$$I = \int |x\rangle dx \langle x|, \quad I = \int |p\rangle dp \langle p| \quad (2.6)$$

## 3. The linear Harmonic oscillator

We will give only a summary of known results on the harmonic oscillator.
*3.1 Schrödinger equation waves functions*
A- The Schrödinger equation of the harmonic oscillator in one dimension is:
$$H\psi(x) = -\frac{\hbar^2}{2m}\frac{d^2\psi(x)}{dx^2} + \frac{m\omega^2}{2}\psi(x) = E\psi(x) \quad (3.1)$$
The energies $E_n = \hbar\omega$ (n+1,2) and the waves functions are [1,2,3]:

$$\psi_n(x) = N_n H_n(q)\, e^{-q^2/2}, \; q = \sqrt{\frac{m\omega}{\hbar}}\, x \quad (3.2)$$

The $H_n(q)$ are the Hermite polynomials of degree *n* and the normalization is:
$$N_n = \left(\frac{1}{2^n n!}\sqrt{\frac{m\omega}{\pi\hbar}}\right)^{1/2} \quad (3.3)$$
The parity of the wave function is given by:
$$\psi_n(-x) = (-1)^n \psi_n(x). \quad (3.4)$$

B- In momentum space $x = i\hbar\partial/\partial p$ the Schrodinger equation becomes:
$$\frac{p^2}{2m}\phi(p) - \hbar^2 m\omega^2 \frac{d^2\phi(p)}{dp^2} = E\phi(p) \quad (3.5)$$
The solution is:
$$\phi_n(p) = \alpha^n C_n H_n(t)e^{-t^2/2} \quad \text{where} \quad t = \frac{p}{\sqrt{m\omega\hbar}} \quad (3.6)$$
With $\alpha^n$ is the phase factor and

$$C_n = \left(\frac{1}{2^n n! \sqrt{m\omega\hbar\pi}}\right)^{1/2} = \frac{N_n}{\sqrt{m\omega}} \qquad (3.7)$$

*3.2 The Harmonic oscillator in Dirac notation*

The Schrodinger equation for the harmonic oscillator is:

$$H\psi(x) = \left(\frac{p^2}{2m} + \frac{m\omega^2}{2}\right)\psi(x) = E\psi(x) \qquad (3.8)$$

This equation is analogous to the product $(z \cdot \bar{z})$, $z \in C$, therefore we put:

$$a = \sqrt{\frac{m\omega}{2\hbar}}q + \frac{1}{\sqrt{2m\omega\hbar}}\frac{d}{dq}, \quad a^+ = \sqrt{\frac{m\omega}{2\hbar}}q - \frac{1}{\sqrt{2m\omega\hbar}}\frac{d}{dq}$$

where $a$, $a^+$ are adjoint operators with:

$$[a, a^+] = 1, \ [a, a] = 0, \ [a^+, a^+] = 0 \qquad (3.9)$$

Comparing $[a, a^+] = 1$ and $[(d/dz), z] = 1$ we simply deduct by analogy and using Dirac notation the orthonormal basis of the harmonic oscillator.

It is well known that the basis of polynomials P(z) is:

$$1, \ z, \ ... \ ..., \ z^n, ... \qquad \text{and} \qquad \frac{d}{dz}1 = 0$$

therefore $\quad |0\rangle, |1\rangle, ..., |n\rangle = \dfrac{a^{+n}}{\sqrt{n!}}|0\rangle, ...$ and $\quad a|0\rangle = 0$.

with $\quad a|n\rangle = n|n-1\rangle, \ a^+|n\rangle = \sqrt{n}|n+1\rangle$

So a and $a^+$ are the ladders operators, $\langle m|n\rangle = \delta_{m,n}$, and $|0\rangle$ is the vacuum state.

The unitary operator of the harmonic oscillator basis is:

$$I = \sum_{n=0}|n\rangle\langle n| \qquad (3.10)$$

Using Dirac transformation [5], the wave function may be written as:

$$u_n(q) = \langle q \| n \rangle = \langle q | \frac{a^{+n}}{\sqrt{n!}} | 0 \rangle \qquad (3.11)$$

And $\quad u_n(q) = \left(\dfrac{\hbar}{m\omega}\right)^{1/4} \psi_n(x), \quad \alpha^n u_n(t) = (m\omega\hbar)^{1/4}\phi_n(p) \qquad (3.12)$

## 4. The generating functions of the harmonic oscillator

We will find the expression of the generating function by a new simple method different from the other methods [1-3] and closely related to Dirac method. This function is the trace of a product of the base $\psi_n(x)$ and the Fock space of analytic functions.

## 4.1 The generating functions

Using Dirac transformation [1-5] and (3.9) we find:

$$\langle q|ae^{za^+}|0\rangle = \frac{1}{\sqrt{2}}(q+\frac{d}{dq})G(z,q) \tag{4.1}$$

Using also (3.9) we find: $\quad \langle q|ae^{za^+}|0\rangle = zG(z,q)$

By comparison of the above expressions we derive:

$$\frac{d}{dq}G(z,q) = (\sqrt{2}z - q)G(z,q) \tag{4.2}$$

The solution of this equation is:

$$G(z,q) = c\exp\{(\sqrt{2}qz - \frac{q^2}{2}) + \varphi(z)\} \tag{4.3}$$

To determine $\varphi(z)$ we use the creation operator, we find:

$$\langle q|a^+e^{za^+}|0\rangle = \frac{1}{\sqrt{2}}(q-\frac{d}{dq})G(z,q) = \frac{\partial}{\partial z}G(z,q) \tag{4.4}$$

WE the help of (4.4) and (4.3) we obtain:

$$\varphi'(z) = -z$$

Therefore the generating function of the oscillator is:

$$G(z,q) = c\exp\{\sqrt{2}qz - \frac{q^2}{2} - \frac{z^2}{2}\}$$

For z = 0, we have $G(0,q) = \langle q|0\rangle = u_0(q)$ and it follows that:

$$G_x(z,q) = \sum_{n=0}^{\infty}\frac{z^n}{\sqrt{n!}}u_n(q) = \pi^{-\frac{1}{4}}\exp\{\sqrt{2}qz - \frac{q^2}{2} - \frac{z^2}{2}\} \tag{4.5}$$

By analogy we find the generating function of the basis $\{\phi_n(p)\}$

$$G_p(z,\alpha t) = \sum_{n=0}^{\infty}\frac{(\alpha z)^n}{\sqrt{n!}}u_n(t) = \pi^{-\frac{1}{4}}\exp\{\sqrt{2}t(\alpha z) - \frac{t^2}{2} - \frac{(\alpha z)^2}{2}\} \tag{4.6}$$

$(\alpha)^n$ is the phase factor of the momentum basis.

## 4.2. The Fock space

In the expression (4.5) the function $f_n(z) = z^n/\sqrt{n!}$ constitutes a basis of analytic Hilbert space known by Fock space and defined by:

$$\langle f_m|f_n\rangle = \iint \overline{f_m(z)}f_n(z)d\mu(z) = \iint \frac{\bar{z}^m}{\sqrt{m!}}\frac{z^n}{\sqrt{n!}}d\mu(z) = \delta_{m,n} \tag{4.7}$$

$d\mu(z)$ is the cylindrical measure or Gaussian measure:

$$d\mu(z) = (1/\pi)e^{-(u^2+v^2)}\,dudv,\, z = u+iv \tag{4.8}$$

## 5. The wave function of momentum

We want to determine the expression $p(x) = \langle x \| p \rangle$ using the generating function of the harmonic oscillator and the orthogonality of Fock space (4.7).
We write:

$$\langle x | p \rangle = \langle x | I | p \rangle = \sum_{n=0} \langle x \| n \rangle \langle n \| p \rangle = \sum_n \psi_n(x) \overline{\phi}_n(p) =$$

$$\sum_{n,m} \frac{1}{\sqrt{\hbar}} u_n(q) \times (\delta_{n,m}) \times \overline{\alpha}^m \overline{u}_m(t) = \frac{1}{\sqrt{\hbar}} \int \sum_{n,m} u_n(q) \frac{z^n}{\sqrt{n!}} \frac{(\overline{\alpha}\,\overline{z})^m}{\sqrt{m!}} \overline{u}_m(t) d\mu(z)$$

$$= \frac{1}{\sqrt{\hbar}} \int G_x(q,z) G_p(t, \overline{\alpha}\,\overline{z}) d\mu(z) \tag{5.1}$$

Using the Gaussian integral

$$\left(\frac{1}{2\pi}\right)^{n/2} \int \prod_{i=1}^n dx_i \exp(-x^t X x + J x) = \frac{1}{\sqrt{(\det(X))}} \exp(\frac{1}{2} J^t X^{-1} J) \tag{5.2}$$

We find

$$X = \begin{pmatrix} 3 + \overline{\alpha}^2 & i - i\overline{\alpha}^2 \\ i - i\overline{\alpha}^2 & 1 - \overline{\alpha}^2 \end{pmatrix},$$

We obtain using (5.2), (3.2) and (3,6):

$$p(x) = \frac{1}{\sqrt{2\pi\hbar}} \exp\left[\frac{q^2 \overline{\alpha}^2 + t^2 \overline{\alpha}^2 - 2tq\overline{\alpha}}{-1 + \overline{\alpha}^2} - \frac{q^2}{2} - \frac{t^2}{2}\right] \tag{5.3}$$

if $2\overline{\alpha}^2 = -1 + \overline{\alpha}^2$, $\overline{\alpha} = \pm i$ we find $\overline{\alpha} = i$ if:

$$p(x) = \frac{1}{\sqrt{2\pi\hbar}} \exp(+\frac{i}{\hbar} xp) \tag{5.4}$$

A simple calculation gives:

$$\frac{1}{\sqrt{2\pi\hbar}} \int \exp(-\frac{i}{\hbar} pq) G_x(z,x) dq = G_p(z,p)$$

We derive after the development of the above expression:

$$\phi_n(p) = \frac{1}{\sqrt{2\pi\hbar}} \int \exp(-\frac{i}{\hbar} xp) \psi_n(x) dx \tag{5.5}$$

We proved that $N_p$ is real and not complex [1,2,3] and we find the same the phase factor of the momentum-space wave function $(-i)^n$ of ref.[1.p:144].

# 6. The Dirac delta function

**A-** With the help of (2,6) we find the expression of the delta function.

$$\langle q'\|q\rangle = \delta(q'-q) = \langle q'|I|q\rangle$$

$$= \int \langle q'\|p\rangle dp \langle p\|q\rangle = \frac{1}{2\pi} \int \exp(+ik(q'-q)) \, dk \tag{6.1}$$

**B-** Using the basis of the oscillator (3.10) and its generating function (4.5) we find a very useful expression of delta function:

$$\langle q'\|q\rangle = \sum_n \langle q'\|n\rangle\langle n\|q\rangle = \sum_n u_n(q')\bar{u}_n(q) \tag{6.2}$$

We deduce also from (3.4) that the delta function is an even function.
Repeating the same method above (5.1), we find

$$\langle q'\|q\rangle = \int G_x(z, q') \overline{G_x(z,q)} \, d\mu(z)$$

Replacing $\overline{G_x(z,q)}$ and $G_x(z,q')$ by the expression (4.5) we obtain

$$\langle q'|q\rangle = \frac{1}{\sqrt{\pi}} \int \exp[-\frac{q'^2+q^2}{2} - \frac{\bar{z}^2+z^2}{2} + \sqrt{2}(q'z+q\bar{z})] d\mu(z) \tag{6.3}$$

The arrangement of this expression gives

$$\langle q'|q\rangle = \frac{1}{\pi\sqrt{\pi}} \exp(-\frac{1}{4}(q-q')^2) \int \exp([-2(u-\frac{\sqrt{2}}{4}(q+q'))^2] + \sqrt{2}iv(q-q'))dudv$$

Using the change of variables and performing the integration we find that

$$\langle q'\|q\rangle = \frac{1}{2\pi} \exp(-\frac{1}{4}(q-q')^2) \int \exp(+ik(q-q')) dk \tag{6.4}$$

Comparing (6.4) and (6.1) and with the help of (2.3) we find:

$$\exp(+\frac{1}{4}(q-q')^2)\langle q'\|q\rangle = \langle q'\|q\rangle = \delta(q'-q)$$

**C-** With the help of the Gauss integral we find that

$$\int_{-\infty}^{+\infty} \delta(q'-q)dq' = \frac{1}{2\pi} \int_{-\infty}^{+\infty}\int_{-\infty}^{+\infty} \exp[-\frac{1}{4}(q-q')^2 + ik(q-q')]dq'dk = 1 \tag{6.5}$$

Therefore we find that the integral of the Dirac delta function is equal to one.

$$\int_{-\infty}^{+\infty} \delta(q'-q)dq' = \frac{1}{2\pi} \int_{-\infty}^{+\infty} \exp[+ik(q-q')]dq'dk = 1 \tag{6.6}$$

Finally, we write:

$$\langle q'\|q\rangle = \delta(q-q') = \frac{1}{2\pi}\int_{-\infty}^{\infty} e^{i(q-q')k} dk \quad \text{and} \quad \int_{-\infty}^{+\infty}\delta(q-q')dq' = 1 \qquad (6.7)$$

So, there is no need to impose (6.6) as constraint [6].

## 7. The Harmonic oscillator propagator

Finally, we note that we obtain by the same simple calculation as above (5.1) the Feynman propagator, [1.p:164, 2.p:119], of the harmonic oscillator:

$$K((x,t),(x',t_0)) = \langle (x,t)|\exp[-iH(t-t_0)/\hbar]|(x',t_0)\rangle \qquad (7.1)$$

We write the propagator in term of the generating function:

$$K((x,t),(x',t_0)) = \left(\frac{m\omega}{\hbar}\right)^{1/2} e^{-i\omega(t-t_0)/2} \int G_x(e^{-i\omega(t-t_0)/2}z,q) G_x(e^{-i\omega(t-t_0)/2}\bar{z},q') d\mu(z) \qquad (7.2)$$

with $q = \sqrt{(m\omega)/\hbar}\, x$, $q' = \sqrt{(m\omega)/\hbar}\, x'$ and $\alpha = \omega(t-t_0)$.
The calculations of (7.2) are carried out quickly by using:

$$(1+e^{-i\alpha}) = 2e^{-i\alpha/2}\cos(\alpha/2),\ (1-e^{-i\alpha}) = 2e^{-i\alpha/2}\sin(\alpha/2)$$
$$\int_{-\infty}^{+\infty} e^{-az^2} dz = \sqrt{\pi/a},\ \text{with}\ \text{Re}(a) > 0.\ \text{and}\quad e^{-i\alpha/2} = 1/\sqrt{e^{-i\alpha}}$$

We obtain the propagator:

$$K((q,t),(q',t_0)) = \sqrt{\frac{m\omega}{2\pi\hbar i \sin\alpha}} \exp[\frac{i}{2\sin\alpha}[(q^2+q'^2)\cos\alpha - 2qq']] \qquad (7.3)$$

Consequently we do not encounter the difficulties of the method proposed by many others [1,2,7]. Therefore our method [4] can be used to calculate the Feynman propagator of charged harmonic oscillator in uniform magnetic field and many other problems.